\documentclass[aps,prb,reprint,superscriptaddress,floatfix]{revtex4-2}

\usepackage[T1]{fontenc}
\usepackage{amsmath,amssymb}
\usepackage{bm}
\usepackage{graphicx}
\usepackage{hyperref}
\hypersetup{colorlinks=true,linkcolor=black,citecolor=blue,urlcolor=blue}
\begin{document}

\title{\texorpdfstring{A rigid commensurate host and a defect that switches:\\
vacancies in the frustrated alternating spin-$(1,\tfrac12)$ Heisenberg chain}{A rigid commensurate host and a defect that switches: vacancies in the frustrated alternating spin-(1, 1/2) Heisenberg chain}}

\author{R.~L.~Silva}
\email{ricardo.l.silva@ufes.br}
\affiliation{Departamento de Ci\^{e}ncias Naturais, Universidade Federal do Esp\'irito Santo,
S\~ao Mateus, ES 29932-540, Brazil}

\author{R.~C.~Silva}
\affiliation{Departamento de Ci\^{e}ncias Naturais, Universidade Federal do Esp\'irito Santo,
S\~ao Mateus, ES 29932-540, Brazil}

\author{A.~R.~Pereira}
\affiliation{Departamento de F\'isica, Universidade Federal de Vi\c{c}osa,
Vi\c{c}osa, MG 36570-000, Brazil}

\begin{abstract}
The alternating spin-$(1,\tfrac12)$ Heisenberg chain with a frustrating
next-nearest-neighbor coupling $J_2$ loses its Lieb--Mattis moment through a
discontinuous transition to a singlet whose nature was left open. Using an
SU(2)-symmetric density-matrix renormalization group, in which the total spin is
a quantum-number label, we identify that phase and follow a single vacancy
through it. The total spin collapses directly from $S=L/4$ to $S=0$, with no
partially polarized window. The singlet phase
is critical: its singlet--triplet gap closes as $1/L$, its entanglement entropy
grows logarithmically with a central charge $c\simeq1.8$, and its modulation
wavevector, extrapolated from $L=80$ to $200$, is commensurate at $3\pi/4$ and
independent of $J_2$ to $5\times10^{-5}$. Removing the central $S=1$ site
switches the ground state of the doped chain from a triplet to a singlet at
$J_2^{\ast}=0.3068(3)$, where the host shows no feature; the triplet is not
bound to the vacancy but spread over one of the two segments it leaves. The
transmission of correlations through the vacancy saturates at a finite value
below $J_2^{\ast}$, $T(\infty)=0.102(15)$ and $0.036(9)$ at $J_2=0.28$ and
$0.30$, and the cut-chain law $T\propto L^{-1}$ is excluded on both sides. We
also show that the longitudinal correlator is not an observable in a triplet
sector.
\end{abstract}

\maketitle

\section{Introduction}
\label{sec:intro}

The alternating spin-$(1,\tfrac12)$ Heisenberg chain is the prototypical
one-dimensional quantum ferrimagnet. Its two inequivalent sublattices,
$S_A=1$ and $S_B=\tfrac12$, place the ground state in the total-spin sector
$S_g=(S_A-S_B)N_{\rm cell}=\tfrac12 N_{\rm cell}$ by the Lieb--Mattis
theorem~\cite{Lieb1962}, so the magnetization per unit cell is pinned at
$m=1/2$ and the spectrum carries a gapless ferromagnetic branch together with a
gapped antiferromagnetic one~\cite{Pati1997,Brehmer1997,Kolezhuk97}. An
antiferromagnetic next-nearest-neighbor coupling $J_2$ frustrates each
sublattice against itself and eventually destroys this moment.

How the moment is destroyed was settled in outline by Ivanov, Richter and
Schollw\"ock~\cite{IvanovRichterSchollwock1998}. Zero-point fluctuations remove
the classical continuous commensurate--incommensurate instability, which for
$S_A/S_B=2$ would occur at $J_2/J_1=0.2$, and replace it by a discontinuous
transition to a singlet at a larger frustration; exact diagonalization of six
unit cells places it at $J_2/J_1\simeq0.229$. Inside the ferrimagnetic phase the
same work locates a disorder point at $J_2/J_1\simeq0.177$, beyond which
real-space correlations acquire an incommensurate component. What the singlet
state above the transition actually is was left open, and stated as such: the
nature of the $S_g=0$ ground states established beyond the ferrimagnetic phase
is named there as one of the important remaining questions. That is the question
we take up. The first-order character of the transition itself is not our
result; it is our starting point, and we reproduce it as a control.

The route to the singlet is not expected to be a single step. In frustrated
mixed-spin systems the generic scenario is an intermediate, partially polarized
ferrimagnetic phase between the Lieb--Mattis ferrimagnet and the nonmagnetic
state. Hida and Takano find such a phase in a mixed-spin chain with frustrated
side chains, carrying an incommensurate spin structure, and observe that this
structure is common to frustration-induced partially polarized ferrimagnets
generally~\cite{HidaTakano2008}. We find that the alternating chain does not
follow this route. With an SU(2)-symmetric density-matrix renormalization group
(DMRG), in which the total spin $S$ labels the sector rather than being inferred
from $\langle\mathbf S^2\rangle$, the ground state at $L=80$ is the Lieb--Mattis
value $S=20$ up to $J_2=0.22$ and $S=0$ from $J_2=0.24$ on. At these two
couplings we computed every sector from $S=0$ to $21$: between $S=0$ and the
Lieb--Mattis value the energy is monotonic in $S$, with its minimum at one end
of the range on one side of the transition and
at the other end on the other side, and what changes across the transition is
the tilt of the whole curve, not the appearance of an interior minimum. The
collapse is direct.

The singlet phase is critical --- its singlet--triplet gap closes as $1/L$ over
$L=80$ to $200$, and its entanglement entropy grows as $\ln L$ with a central
charge $c\simeq1.8$ --- and, within our resolution, rigid. Measured
through the SU(2)-invariant correlator
$\langle\mathbf S_i\!\cdot\!\mathbf S_j\rangle$ and extrapolated in $1/L$, the
modulation wavevector is commensurate, $q=3\pi/4$, and does not move with
$J_2$; at fixed $L$ it drifts monotonically, but the drift is a finite-size
correction whose amplitude shrinks with $J_2$ while the extrapolated value
stays put; the coupling at which that correction changes sign is itself a
finite-size quantity, moving from $0.3076(43)$ measured on $L=80,120$ to
$0.3263(18)$ on $L=120,160$ and back below $0.30$ on $L=160,200$, where the
correction is no longer resolved (Sec.~\ref{sec:pitch}).

That rigidity is what makes the defect problem sharp. A single impurity in a
one-dimensional interacting system is a singular perturbation: Kane and Fisher
showed that for $K<1$ an arbitrarily weak barrier grows under renormalization
until it cuts the chain~\cite{KaneFisher1992a,KaneFisher1992b}, while internal
degrees of freedom of the defect can open resonant channels and change that
fate~\cite{KaneFisher1992c,FurusakiNagaosa1994,EggertAffleck1992}. For a
single $S=1$ site removed from this chain, a preliminary study of ours with a
U(1)-symmetric code and the longitudinal correlator (unpublished; the U(1)
runs below) pointed to two behaviors: below a critical frustration, a vacancy
that binds a spin-1 moment and transmits correlations with a scale-invariant
plateau; above it, a vacancy that cuts the chain. We show here
(Sec.~\ref{sec:invariant}) that in the triplet sector the longitudinal
correlator depends on the magnetic quantum number $m$ of the member probed
and is therefore not an observable, so the transmission has to be rebuilt
from the invariant. Rebuilt (Sec.~\ref{sec:defect}), it separates the two
sides of a coupling $J_2^{\ast}$ by the sign of its finite-size correction:
below $J_2^{\ast}$ it saturates at a finite value, above it keeps falling,
and on neither side does it follow the cut-chain law.
The spin-1 moment, however, does not survive: below $J_2^{\ast}$ the doped
chain is a triplet, but the triplet is a magnon of the host confined to one
segment, not a moment bound to the vacancy, and $J_2^{\ast}$ itself depends on
where the vacancy sits.

The two transitions are, on the evidence assembled here, unrelated: the defect
changes its nature at $J_2^{\ast}$ while the host passes through that coupling
without a feature in its thermodynamic pitch. A defect transition in a rigid
commensurate host is the
central result of this work.

The system is not hypothetical. Bimetallic chains realize alternating
$(S,S')=(1,\tfrac12)$ Heisenberg ferrimagnets~\cite{Kahn1987,Hagiwara1998}, and
atom-by-atom assembly of mixed-spin structures has reached the point where such
chains can be built to specification~\cite{Huang25,Mishra21}, which makes the
tuning of $J_2$ a less remote prospect than it once was.

The paper is organized as follows. Section~\ref{sec:model} defines the model and
the vacancy convention. Section~\ref{sec:method} describes the SU(2) DMRG, its
validation, and the invariant observables. Section~\ref{sec:bulk} gives the
total-spin map and establishes the direct collapse. Section~\ref{sec:pitch}
settles the commensurability of the pitch. Section~\ref{sec:defect} presents
the vacancy, and Section~\ref{sec:discussion} discusses the separation of bulk
and defect physics.

\section{Model and lattice}
\label{sec:model}

We consider
\begin{equation}
H = J_1 \sum_{|i-j|=1} \mathbf{S}_i\cdot\mathbf{S}_j
  + J_2 \sum_{|i-j|=2} \mathbf{S}_i\cdot\mathbf{S}_j ,
\label{eq:H}
\end{equation}
on an open chain of $L$ sites, with $S=1$ on odd sites (sublattice $A$) and
$S=\tfrac12$ on even sites (sublattice $B$), so that $N_{\rm cell}=L/2$. Both
couplings are antiferromagnetic and $J_1=1$ sets the energy scale. The sums run
over pairs at distance one and two \emph{in the original site index}; this
phrasing is what makes the vacancy convention unambiguous.

\emph{Vacancy convention.} A vacancy is the removal of one site from the list of
sites, with Eq.~(\ref{eq:H}) still evaluated by the original distance rule.
Removing site $v$ leaves its two neighbors $v\pm1$ at original distance two, so
the $J_2$ bond between them survives: the hole is bridged by a single $J_2$
coupling, and both sublattices are preserved on either side. This is not a side
effect of bookkeeping; it is the physics of a missing magnetic ion in a chain
whose exchange paths are set by geometry. We use three lattices: \texttt{clean}
(all $L$ sites), \texttt{vacS1} (the central $S=1$ site removed; integer total
spin), and \texttt{vacS12} (the central $S=\tfrac12$ site removed; half-integer
total spin). Removal of the spin-$\tfrac12$ site dresses the defect with a
delocalized magnon and does not measure transmission through the defect; it is
kept as a control for the sector analysis only: with an $S=\tfrac12$ site
removed the chain has half-integer total spin and its correlations across the
defect are dominated by the $\pi$ component, so the incommensurate transmission
of Sec.~\ref{sec:estimators} is not a meaningful probe there and is not
reported.

\section{Method: SU(2) DMRG and invariant observables}
\label{sec:method}

\subsection{SU(2)-symmetric DMRG}

Ground states are obtained with a two-site DMRG~\cite{White1992,Schollwock2011}
that exploits the full SU(2) symmetry~\cite{McCulloch2002,Singh2010},
implemented with the SU(2)-symmetric tensor libraries
\textsc{TensorKit} and \textsc{MPSKit}~\cite{TensorKit,MPSKit} in Julia. The
total spin $S$ of the state is a label of the symmetry sector in which the
sweep is carried out, not a quantity inferred afterwards from
$\langle\mathbf S^2\rangle$. Truncation is by a cap on the total bond dimension, counted as in a
U(1) or dense code (each SU(2) multiplet of spin $s$ contributes $2s+1$
states); we set it to $2400$, $3600$, $4800$ and $6000$ for $L=80$, $120$,
$160$ and $200$, scaling with $L$. The number of SU(2) multiplets actually
stored on the most entangled bond is about one quarter of that ($\simeq610$
and $\simeq910$ for $L=80$ and $120$, $\simeq1200$ for $L=160$,
$\simeq1500$ for $L=200$), which is the compression the symmetry buys. Sector
energies used to locate the ground state are converged to a
tolerance of $10^{-8}$ in $E_0$ with at most $100$ sweeps, sufficient to
separate sectors whose gaps are $\gtrsim10^{-3}$; the states used for
correlators are converged to $10^{-12}$ with at most $400$ sweeps. Six
states were converged to $10^{-10}$ with $200$ sweeps instead: three that
did not reach $10^{-12}$ within a wall-time cap (all within $0.002$ of a
level crossing, where the sweep stalls) --- the clean chain at $L=160$,
$J_2=0.31$, and the doped chains \texttt{vacS12} at $L=80$, $J_2=0.26$ and
\texttt{vacS1} at $L=120$, $J_2=0.30$ --- and, at $L=200$, the doped chain
\texttt{vacS1} at $J_2=0.28$ and $0.30$ (in the $S=1$ sector, whose sweep at
$10^{-12}$ had not converged after $19$ hours) and the clean chain at
$J_2=0.30$. On a state that converges either way, relaxing the tolerance
changes $E_0$ by $10^{-14}$, the pitch by $6\times10^{-10}$ and the
correlators by $4\times10^{-10}$; at $L=200$, $J_2=0.28$, the $10^{-10}$
ground-state energy agrees with the $10^{-8}$ sector probe to
$2\times10^{-11}$.

The singlet--triplet gaps of Sec.~\ref{sec:pitch} and the binding energies of
Sec.~\ref{sec:defect} use a second set of states, converged at bond dimension
$3600$ for $L\ge120$ ($2400$ at $L=80$), to $10^{-10}$ with at most $200$
sweeps, and then polished by five sweeps of a one-site DMRG with a tight local
eigensolver. The polishing addresses a limitation we found when validating
against exact diagonalization at $L=12$, where the bond dimension spans the
full space: the two-site sweep stops with an energy variance
$\langle H^2\rangle-\langle H\rangle^2\sim10^{-8}$, which costs $10^{-4}$ in
local observables, and the one-site sweeps bring it to $10^{-13}$. At $L\ge80$
the residual variance after polishing ($10^{-11}$ to $10^{-7}$) is set by
truncation instead. For one state, the clean chain at $L=200$, $J_2=0.28$,
$S=1$, the variance and the polishing step exceed the memory of the machine
used and were omitted; on the other $L=200$ states the polishing moved $E_0$
by at most $2\times10^{-8}$. Where states at the larger bond dimensions of the
previous paragraph also exist, the energies at $3600$ differ from them by at
most $3\times10^{-9}$ at $L=160$ and $6\times10^{-8}$ at $L=200$, five orders
of magnitude below the gaps.

\emph{Validation.} The engine reproduces exact diagonalization in 36 cases
($L=8,10,12$; the three lattices; four values of $J_2$) to $10^{-9}$ in $E_0$
and exactly in the total spin. The exact diagonalization is itself anchored to a
theorem: at $J_2=0$ the chain is bipartite with unequal sublattices and
Lieb--Mattis fixes $S=|S_A-S_B|$; the predicted values $2,\,2.5,\,3$
(\texttt{clean}), $1,\,1.5,\,2$ (\texttt{vacS1}) and $2.5,\,3,\,3.5$
(\texttt{vacS12}) for $L=8,10,12$ are reproduced in all nine cases.

\emph{Sector search.} Locating the ground-state sector by a local search around
the value found at the previous $J_2$ fails at a first-order transition, and
this failure is instructive: at $L=80$, $J_2=0.26$, such a search returns
$S=17$ with $E=-47.8575$ while $S=0$ has $E=-48.4046$, lower by $0.547$. The
ground state jumps from $S=L/4$ to $S=0$ without passing through the
intermediate sectors, so the low sectors must always be included in the
comparison. All maps reported below do so.

\subsection{The invariant correlator}
\label{sec:invariant}

We measure the SU(2) invariant $C(i,j)=\langle\mathbf S_i\!\cdot\!\mathbf S_j\rangle$
rather than the longitudinal $C^{zz}(i,j)=\langle S^z_iS^z_j\rangle$ used in
the U(1) runs. The difference is not one of normalization. Exact
diagonalization at $L=12$ shows that in a singlet the identity
$\langle\mathbf S_i\!\cdot\!\mathbf S_j\rangle=3\langle S^z_iS^z_j\rangle$ holds
to a residual of $4\times10^{-15}$, so for the clean chain in its singlet phase
the pitch and any ratio of correlators are unchanged by the substitution. In an
$S=1$ multiplet the identity fails by $0.3$--$0.76$, and $C^{zz}$ differs
between the $m=-1$ and $m=0$ members by up to $0.32$ while the invariant differs
by $7\times10^{-14}$. The longitudinal correlator in a triplet is therefore a
property of the member probed, not of the state, and the transmission through a
vacancy whose doped chain is a triplet has to be built from the invariant.

A second point follows from the symmetry. In an SU(2) eigenstate
$\langle\mathbf S_i\rangle=0$ identically, so there is no disconnected part to
subtract. Ferrimagnetic order, where present, appears as a non-decaying,
alternating $C(i,j)$, i.e.\ as weight at $k=\pi$; the incommensurate peak is a
separate feature, and the pitch search is restricted to a window that excludes
$\pi$.

\subsection{Estimators}
\label{sec:estimators}

\emph{Pitch.} For the clean chain we form the bulk-averaged correlator
$C(r)$ from pairs $(i,i+r)$ with $12$ reference sites $i$ in the central half
of the chain and $r\le0.45L$, and take the peak of its Hann-windowed Fourier
transform in $k\in[0.55\pi,0.95\pi]$, refined by three-point parabolic
interpolation. The estimator, windows and interpolation are identical to those
used with $C^{zz}$ in our U(1) runs, so that
the comparison with the U(1) numbers is one of identical estimator on a
different observable. The thermodynamic pitch is obtained by linear
extrapolation in $1/L$; the U(1) runs, with four or five sizes
($L=80$--$256$), showed the $1/L$ form to be preferred over $1/L^2$ by a
factor $11$--$25$ in residual.

\emph{Transmission.} The transmission of the modulated correlations across the
vacancy at $v$ is the ratio of Fourier amplitudes of the symmetric-pair
correlator $C(v-r,v+r)$, defect over clean, evaluated at the clean pitch
$q_{\rm IC}$ with a Hann window over $r/L\in[0.08,0.22]$ and $[0.10,0.25]$
(the spread sets the systematic error):
\begin{equation}
T_{\rm IC}(L)=\frac{|F_{\rm def}(q_{\rm IC})|}{|F_{\rm clean}(q_{\rm IC})|}.
\label{eq:T}
\end{equation}
The clean chain is sampled at the same center as the defect it normalizes,
which for \texttt{vacS1} is an odd site and for \texttt{vacS12} an even one.
Power counting at the cut-chain fixed point, to first order in the tunneling
between the half chains, gives $T_{\rm IC}\propto L^{1-2\eta}$, with $\eta$ the
decay exponent of the bulk spin correlator; the measured $\eta=1.0(1)$
(Sec.~\ref{sec:pitch}) makes this $T_{\rm IC}\propto L^{-1}$
(Appendix~\ref{app:scaling}). The criterion, fixed
before the larger sizes were run, is that pairwise slopes $b(L_1\!\to\!L_2)$
rise toward $1$ if the defect flows to
the cut chain, and stay near zero on a plateau.

\section{Total spin of the clean chain: direct collapse}
\label{sec:bulk}

\begin{table}[tbp]
\begin{center}
\begin{tabular}{cccc}
\hline\hline
$J_2$ & $S$ & $E_0$ & gap \\
\hline
0.00 & 20 & $-57.687761$ & $2.6\times10^{-3}$ \\
0.10 & 20 & $-53.708120$ & $1.4\times10^{-3}$ \\
0.20 & 20 & $-49.902833$ & $3.3\times10^{-4}$ \\
0.22 & 20 & $-49.168487$ & $1.1\times10^{-4}$ \\
0.24 & 0  & $-48.571546$ & $4.3\times10^{-4}$ \\
0.26 & 0  & $-48.404611$ & $2.8\times10^{-3}$ \\
0.29 & 0  & $-48.302448$ & $1.2\times10^{-2}$ \\
0.31 & 0  & $-48.284625$ & $1.6\times10^{-2}$ \\
0.33 & 0  & $-48.309441$ & $1.7\times10^{-2}$ \\
0.36 & 0  & $-48.431683$ & $1.9\times10^{-2}$ \\
0.40 & 0  & $-48.773046$ & $2.2\times10^{-2}$ \\
\hline\hline
\end{tabular}
\end{center}
\caption{Ground-state total spin $S$, energy $E_0$ and gap to the nearest
competing sector of the clean chain at $L=80$, from the SU(2) sector search
(bond dimension 2400). The Lieb--Mattis value is $S=L/4=20$. At $J_2=0.22$ and
$0.24$, which bracket the transition, every sector from $S=0$ to $21$ was
computed (bond dimension 800 for the sweep, 2400 for the winner and its
neighbors); the gap column at these two couplings is the distance to the nearest
sector, $S=19$ and $S=1$ respectively.}
\label{tab:Smap}
\end{table}

Table~\ref{tab:Smap} gives the ground-state sector of the clean chain at
$L=80$. The Lieb--Mattis value $S=20$ persists up to $J_2=0.22$, and from
$J_2=0.24$ onward the ground state is the singlet. At these two couplings,
which bracket the transition, we did not rely on a local search: every sector
from $S=0$ to $21$ was computed. At $J_2=0.22$ the energy decreases
monotonically with $S$, from $-49.1655$ at $S=0$ to $-49.1685$ at $S=20$, in
steps of order $10^{-4}$; at $J_2=0.24$ it increases monotonically with $S$,
from $-48.5715$ at $S=0$ to $-48.4447$ at $S=20$, in steps of order $10^{-2}$.
The sector $S=21$, beyond the Lieb--Mattis value, lies more than one unit of
$J_1$ above the ground state at both couplings. Neither curve has an interior
minimum. What changes between the two couplings
is the tilt of the whole $E(S)$ curve, whose minimum jumps from one end of the
range to the other; a partially polarized phase would instead require an
interior minimum to develop, and none does. At couplings above $0.26$ the
competing sectors are higher in energy by at least the gaps listed, and at
$0.26$ the $S=17$ state found by a local search is $0.55$ above the singlet.
The collapse from $S=L/4$ to $S=0$ is direct, with no partially polarized
window of the kind that the frustrated mixed-spin scenario of
Ref.~\cite{HidaTakano2008} would predict.

Two remarks on what this table does and does not establish. First, the
transition is bracketed between $0.22$ and $0.24$; the finer value
$J_2^{c1}=0.233(3)$ from the $S^z$-resolved U(1) runs (Maxwell construction and
lever rule, Appendix~\ref{app:maxwell}) falls inside this bracket and is
consistent with the six-cell estimate $0.229$ of
Ref.~\cite{IvanovRichterSchollwock1998}. Second, the singlet just above a
first-order transition is a hard state for the DMRG: at $J_2=0.24$ the low
sectors took several hours each even at bond dimension 800. The sector
ordering, however, is decided by margins of $10^{-4}$ and more, far above any
truncation effect: between bond dimensions 800 and 2400 the $S=20$ energy at
$J_2=0.22$ moves by $6\times10^{-9}$ and the $S=0$ energy at $J_2=0.24$ by
$8\times10^{-6}$.

The gap column is the distance to the nearest sector, not the spin gap of the
singlet phase; it is included because it is the quantity the sector search
resolves, and it sets the tolerance a sector certification needs.

\section{The pitch of the singlet phase}
\label{sec:pitch}

\begin{table*}[t]
\begin{center}
\begin{tabular}{ccccccc}
\hline\hline
$J_2$ & $L=80$ & $L=120$ & $L=160$ & $L=200$ & $q(\infty)/\pi$ & $L=80$, $C^{zz}$ \\
\hline
0.26  & 0.75157 & 0.75139 & 0.75152 & ---     & 0.75137 & 0.75164 \\
0.27  & 0.75065 & 0.75045 & 0.75034 & ---     & 0.75003 & \\
0.275 & 0.75048 & 0.75031 & 0.75019 & ---     & 0.74992 & \\
0.28  & 0.75036 & 0.75022 & 0.75011 & 0.75010 & 0.74990 & 0.75034 \\
0.30  & 0.75008 & 0.75004 & 0.74996 & 0.74998 & 0.74989$^{\rm a}$ & 0.75010 \\
0.305 & 0.75002 & 0.75001 & 0.74993 & ---     & 0.74987 & \\
0.3075 & 0.74999 & 0.74999 & 0.74992 & ---    & 0.74988 & \\
0.31  & 0.74996 & 0.74997 & 0.74990 & ---     & 0.74988 & \\
0.3125 & 0.74994 & 0.74997 & 0.74989 & ---    & 0.74989 & \\
0.32  & 0.74981 & 0.74988 & 0.74985 & 0.74990 & 0.74994 & 0.74984 \\
0.35  & 0.74909 & 0.74944 & 0.74958 & ---     & 0.75009 & \\
\hline\hline
\end{tabular}
\end{center}
\par\smallskip\noindent{\footnotesize $^{\rm a}$ Sizes $80$--$200$; including $L=40$ ($q/\pi=0.75207$) in the
fit gives $0.74919$ with a residual $2.5\times10^{-4}$, i.e.\ $L=40$ is
outside the asymptotic regime.}
\caption{Modulation wavevector $q/\pi$ of the clean chain in the singlet
sector, from the invariant correlator, at $L=80$, $120$, $160$ and $200$
(bond dimension 2400, 3600, 4800, 6000), and its linear extrapolation in
$1/L$; the fit uses all sizes available at that $J_2$. Every coupling has
three sizes; $0.28$, $0.30$ and $0.32$ have four. Last column: the $C^{zz}$
result of the U(1) runs at $L=80$ where available.}
\label{tab:pitch}
\end{table*}

\emph{Agreement with the longitudinal estimator.} Where the clean chain is a
singlet the invariant and longitudinal estimators must agree, and they do:
at $L=80$ and $J_2=0.26,\,0.28,\,0.30,\,0.32$ the new pitch differs from the
$C^{zz}$ value by at most $6\times10^{-5}\pi$ (Table~\ref{tab:pitch}, last
column), a test of the engine rather than a new result.

\emph{Finite-size drift and extrapolation.} At fixed $L$ the pitch decreases
monotonically with $J_2$, from $0.7516\pi$ at $J_2=0.26$ to $0.7491\pi$ at
$0.35$ ($L=80$). The drift is a finite-size correction whose amplitude shrinks with
$J_2$, and whose sign depends on which pair of sizes measures it. Taking the
slope $b=\mathrm{d}q/\mathrm{d}(1/L)$ from the $L=80,120$ pair at every
coupling, so that one estimator is used throughout, $b=+0.033$ at $J_2=0.28$,
$+0.003$ at $0.305$, $-0.003$ at $0.31$, $-0.018$ at $0.32$ and $-0.086$ at
$0.35$: it passes through zero at
\begin{equation}
  J_2^{b=0}(80,120)=0.3076\pm0.0043 ,
  \label{eq:bzero}
\end{equation}
close to where the defect changes sector. The choice of Fourier window shifts
$q$ at the two sizes by the same amount and cancels in $b$; what does not
cancel is how much of the tail of $r$ enters the transform, and that sets the
uncertainty, $\pm0.004$ near the sign change, which is set by the per-point
error and not by the spacing in $J_2$.

\emph{The sign change is a property of the pair, not of the limit.} On the
$L=120,160$ pair, available at every coupling, the same estimator gives $b$
\emph{positive throughout} the window: $+0.055$ at $J_2=0.27$,
$+0.057$ at $0.28$, $+0.042$ at $0.30$, $+0.035$ at $0.3075$, $+0.032$ at
$0.31$, $+0.028$ at $0.3125$ and $+0.017$ at $0.32$, each with an uncertainty
of $\pm0.003$ from the same tail-of-$r$ systematic. This pair crosses zero
only between $0.32$ and $0.35$, at $J_2^{b=0}(120,160)=0.3263\pm0.0018$,
$9\sigma$ away from $J_2^{\ast}(L{=}80)$ and $0.019$ above the crossing of the smaller
pair; the coupling $0.26$, next to the first-order boundary, is the only one
where it is negative below $0.35$. The two pairs disagree about the sign of
the correction at every coupling between $0.3076$ and $0.32$, by ten times their common
uncertainty, so $q(L)$ does not follow a pure $1/L$ over these sizes. On the
largest pair, $L=160,200$, available at $J_2=0.28$, $0.30$ and $0.32$, the
pitch has stopped moving: $q(160)$ and $q(200)$ differ by $1$, $2$ and
$5\times10^{-5}$, at the level of the fit residuals, and the slope on that
pair ($+0.008$, $-0.016$, $-0.040$) is not resolved from zero at the first
two couplings. The coupling where the finite-size correction vanishes is
therefore itself a finite-size quantity that moves with the pair ---
$0.3076$ on $(80,120)$, $0.3263$ on $(120,160)$, below $0.30$ on
$(160,200)$ --- and the proximity of $J_2^{b=0}(80,120)$ to
$J_2^{\ast}(L{=}80)$ is a coincidence of one pair of sizes, not a relation
between two limits.

\begin{figure}[tbp]
\centering
\includegraphics[width=\linewidth]{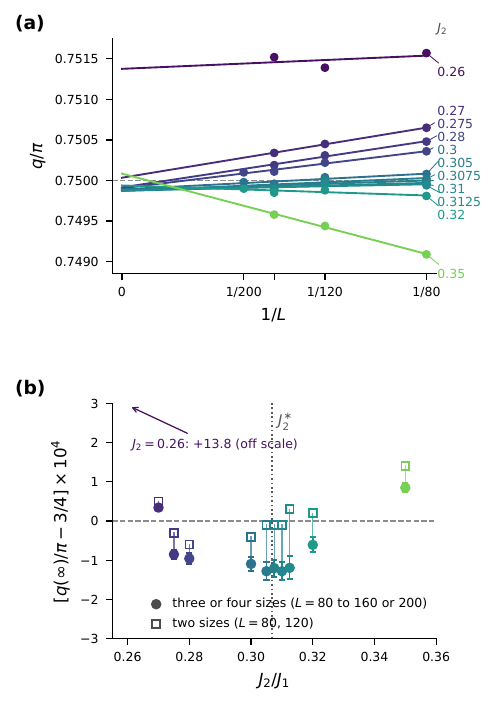}
\caption{Pitch of the clean chain in the singlet phase. (a) $q(L)/\pi$ against
$1/L$ for each $J_2$ (labels at the right), with the linear extrapolation;
the dashed line is $3/4$. The slope shrinks with $J_2$; measured on the $L=80,120$ pair it changes
sign at $J_2=0.3076(43)$, Eq.~(\ref{eq:bzero}), but on the $L=120,160$ pair it
stays positive across the whole window --- the crossing is a property of the
pair of sizes, not of the limit; on $L=160,200$ the pitch has stopped moving.
(b) Extrapolated pitch, $q(\infty)/\pi-3/4$, against $J_2$; filled circles use
three or four sizes ($L=80$ to $160$ or $200$) with the rms fit residual as
error bar; open squares use two sizes ($L=80$, $120$) and are biased upward by
$\simeq10^{-4}$, as the vertical connectors at the completed couplings show.
The dotted line marks the defect transition $J_2^{\ast}$. On the couplings
measured alike the pitch does not move across $J_2^{\ast}$.}
\label{fig:pitch}
\end{figure}

\emph{Two sizes are not enough.} How many sizes enter the fit matters at the
level of the effect we are after. Dropping $L=160$ and extrapolating from $L=80$
and $120$ alone raises
$q(\infty)/\pi$, for example by $0.6$, $1.2$ and $1.0\times10^{-4}$ at $J_2=0.28$, $0.305$
and $0.32$. The two-size estimator is biased upward by about
$10^{-4}$ --- the same size as the departure from $3/4$ it appears to resolve.
Every row of Table~\ref{tab:pitch} therefore uses at least three sizes; the
two-size values are drawn as open squares in
Fig.~\ref{fig:pitch}b, where the bias is visible as the offset between the two
symbols at each coupling.

\emph{The thermodynamic pitch does not move with $J_2$.} On five of the
couplings measured with the same estimator, $q(\infty)/\pi$ is $0.75003$,
$0.74990$, $0.74987$, $0.74988$ and $0.74994$ at $J_2=0.27$, $0.28$, $0.305$,
$0.31$ and $0.32$ (four sizes at $0.28$ and $0.32$, three elsewhere): equal
to $1.6\times10^{-4}$ over the whole interval and to $5\times10^{-5}$ over
$[0.28,0.32]$, where the three couplings with four sizes give $0.74990$,
$0.74989$ and $0.74994$ at $0.28$, $0.30$ and $0.32$, on both sides of the
defect transition $J_2^{\ast}$. The pitch of the host is thus rigid, and in
particular shows no feature where the ground-state sector of the doped chain
switches. What changes near $J_2\simeq0.31$ is only the direction from which
$q(L)$ approaches that limit on the $L=80,120$ pair, where
$J_2^{b=0}(80,120)=0.3076(43)$ sits $0.6\sigma$ from
$J_2^{\ast}(L{=}80)=0.3102$; on $L=120,160$ there is no sign change anywhere
in the window. Restricting the correlator sample to the interior of the
chain moves $J_2^{b=0}(80,120)$ between $0.306$ and $0.318$ without a trend,
and the defect crossing itself moves by $1.3\times10^{-3}$ between $L=80$
and $120$. The host registers $J_2^{\ast}$ in nothing that survives to the
thermodynamic limit.

\emph{What is not resolved.} Three sizes fit the linear form to an rms residual
of $\le2.3\times10^{-5}$, but that residual is not small compared with the
total variation of $q(L)$ over the same sizes, and the pair-dependence of $b$
above measures the same curvature more sharply: three points cannot separate
$1/L$ from $1/L+1/L^{2}$, and a quadratic through them is exact by
construction and returns $q(\infty)/\pi=0.7496$, $0.7494$ and $0.7495$. The
absolute offset of the thermodynamic pitch from $3/4$ is therefore not resolved
by the present data --- it is $0$, $-1\times10^{-4}$ or $-5\times10^{-4}$
according to the fit form. What is resolved, because one estimator is used
throughout, is that this offset does not depend on $J_2$. We take the
modulation to be commensurate at $3\pi/4$ and rigid. A fourth size at the
three couplings $0.28$, $0.30$ and $0.32$ narrows the linear-fit offset to
$-1.0$, $-1.1$ and $-0.6\times10^{-4}$ with residuals $\le2\times10^{-5}$,
but the change between $L=160$ and $200$ is of the same order as those
residuals, so the offset remains a bound; the four sizes fix the plateau and
its independence of $J_2$, which is what the conclusions rest on.

The residual at $J_2=0.26$ is the only resolved departure. With three sizes
the extrapolated pitch there is $q(\infty)/\pi=0.75137$, which is
$+1.4\times10^{-3}$ above $3/4$, with a fit
residual ($6.8\times10^{-5}$) three times larger than at the other couplings,
so that the $1/L$ form is already strained there. It lies next to the
first-order boundary $J_2^{c1}$; whether it is a genuine incommensurability
or the onset of the non-asymptotic regime seen at $L=40$ is not decided by the
present data.

\emph{Below $J_2=0.26$.} Immediately above the transition, at $J_2=0.24$
and $0.25$, the pitch is not reported: the ground-state sector at $0.24$ is
certified as the singlet (Table~\ref{tab:Smap}), but the pitch has not been
measured there with the SU(2) engine, and we do not quote pitches measured in
an inferred sector.

\emph{The singlet phase is gapless.} The rigidity of the pitch does not by
itself say whether the phase has a gap. We measured the singlet--triplet gap
of the clean chain, $\Delta_{\rm ST}=E_0(S{=}1)-E_0(S{=}0)$, at $J_2=0.28$
and $0.30$ on the four sizes (Table~\ref{tab:gapST}). At both couplings
$L\Delta_{\rm ST}$ is nearly constant --- $0.730$, $0.719$, $0.707$, $0.696$ at
$0.28$ and $1.189$, $1.215$, $1.229$, $1.237$ at $0.30$, for $L=80$ to $200$
--- and the extrapolation $\Delta_{\rm ST}=\Delta_\infty+b/L$ returns
$\Delta_\infty=-0.29\times10^{-3}$ and $+0.45\times10^{-3}$. Adding a $1/L^2$
term, or dropping $L=80$, keeps the intercepts within $0.3\times10^{-3}$ of
zero at $0.28$ and moves them to $+0.01\times10^{-3}$ and $+0.28\times10^{-3}$
at $0.30$. The gap therefore closes as $1/L$ to within a few times $10^{-4}$,
a twentieth of its value at $L=200$; the residual drift of $L\Delta_{\rm ST}$,
a few percent over a factor $2.5$ in $L$ and of opposite sign at the two
couplings, is what a marginal logarithmic correction would produce. This is
the direct evidence that the singlet phase is critical, which the pitch alone
does not supply.

\emph{Central charge and correlation exponent.} The entanglement entropy of the
clean chain identifies the critical theory more closely. For an open chain the
von Neumann entropy of the first $l$ sites follows
$S(l)=\frac{c}{6}\ln\big[\frac{2L}{\pi}\sin\frac{\pi l}{L}\big]+{\rm const}$
up to oscillating terms. We fit each residue class $l\bmod8$ separately over
$0.2L\le l\le0.8L$, which removes the period-eight oscillation, and average
the eight slopes, which scatter by $\pm0.2$--$0.3$ about their mean. At
$J_2=0.28$ and $0.30$ this gives $c=1.73$, $1.77$, $1.80$, $1.82$ and $1.72$,
$1.76$, $1.79$, $1.81$ for $L=80$ to $200$ (Table~\ref{tab:cc}); the entropy
averaged over one period at the center grows as $(c/6)\ln L$ across the four
sizes with $c=1.82$ at both couplings, and the windows $[0.1,0.9]L$ and
$[0.3,0.7]L$ move $c$ by less than $0.05$. The central charge is therefore
well above the value $c=1$ of a single SU(2)$_1$ chain and still drifts
slowly upward with $L$; these sizes place its limit between $1.8$ and $2$. The
spin correlations, by contrast, decay with the exponent of a single SU(2)$_1$
channel: the envelope of the bulk correlator $C(r)$, fitted class by class
($r\bmod8$) over $0.05L\le r\le0.35L$ against the chord distance
$\frac{2L}{\pi}\sin\frac{\pi r}{2L}$, gives $\eta=1.04$ at $J_2=0.28$ and
$0.99$ at $0.30$ for $L=200$; from $L=80$ to $200$ the value changes by at
most $0.02$, and by at most $0.04$ when the plain distance $r$ or the window
$[0.1,0.3]L$ is used. A spin
channel with $\eta=1$ alongside a central charge near $2$ points to a second
gapless mode that does not carry the leading spin correlation; identifying it
is beyond the scope of this work. For the transmission through a vacancy the
exponent that enters is $\eta$ (Appendix~\ref{app:scaling}).

\begin{table}[tbp]
\begin{center}
\begin{tabular}{cccccc}
\hline\hline
$J_2$ & $L=80$ & $120$ & $160$ & $200$ & $\Delta_\infty$ \\
\hline
0.28 & $9.129$ & $5.991$ & $4.418$ & $3.480$ & $-0.29$ \\
0.30 & $14.868$ & $10.129$ & $7.681$ & $6.185$ & $+0.45$ \\
\hline\hline
\end{tabular}
\end{center}
\caption{Singlet--triplet gap of the clean chain, $\Delta_{\rm ST}\times10^{3}$,
and its linear extrapolation in $1/L$. States at bond dimension $2400$
($L=80$) and $3600$ ($L\ge120$); see Sec.~\ref{sec:method} for convergence and
truncation.}
\label{tab:gapST}
\end{table}

\begin{table}[tbp]
\begin{center}
\begin{tabular}{cccccc}
\hline\hline
$J_2$ & $L=80$ & $120$ & $160$ & $200$ & from $S_{\rm center}(L)$ \\
\hline
0.28 & $1.73$ & $1.77$ & $1.80$ & $1.82$ & $1.82$ \\
0.30 & $1.72$ & $1.76$ & $1.79$ & $1.81$ & $1.82$ \\
\hline\hline
\end{tabular}
\end{center}
\caption{Central charge of the clean chain from the entanglement entropy of
the singlet ground state: Calabrese--Cardy fit for an open chain at each size
(window $0.2L\le l\le0.8L$, mean over the eight residue classes $l\bmod8$),
and from the growth of the entropy at the center, averaged over one period,
across the four sizes. States as in Table~\ref{tab:gapST}. A single SU(2)$_1$
chain has $c=1$.}
\label{tab:cc}
\end{table}

\section{The vacancy}
\label{sec:defect}

\subsection{Ground-state sector of the doped chain and \texorpdfstring{$J_2^{\ast}$}{J2*}}

In a U(1) code the total spin of the doped chain can only be inferred, from
the splitting of the two lowest $S^z$ levels and from
$\langle\mathbf S^2\rangle$. Here
the sector is a label: for each $J_2$ we converge \texttt{vacS1} in $S=0$ and
$S=1$ (and \texttt{vacS12} in $S=\tfrac12$ and $\tfrac32$) and the ground
state is the lower of the two.
Table~\ref{tab:sectors} collects the two sector energies at $L=80$.

\begin{table*}[t]
\begin{center}
\begin{tabular}{cccccc}
\hline\hline
$J_2$ & $E_0(S{=}1)$ & $E_0(S{=}0)$ & $\Delta\times10^{3}$ & $\Delta_{\rm ST}\times10^{3}$ & $E_b\times10^{3}$ \\
\hline
\multicolumn{6}{c}{\texttt{vacS1}} \\
\hline
0.26 & $-47.899177$ & $-47.893187$ & $-5.99$ & $2.83$ & $8.8$ \\
0.28 & $-47.811167$ & $-47.801475$ & $-9.69$ & $9.13$ & $18.8$ \\
0.30 & $-47.760540$ & $-47.757218$ & $-3.32$ & $14.87$ & $18.2$ \\
0.31 & $-47.749887$ & $-47.749822$ & $-0.065$ & $15.77$ & $15.8$ \\
0.315 & $-47.748356$ & $-47.749922$ & $+1.57$ & $16.16$ & $14.6$ \\
0.32 & $-47.749405$ & $-47.752607$ & $+3.20$ & $16.53$ & $13.3$ \\
\hline
 & $E_0(S{=}\tfrac32)$ & $E_0(S{=}\tfrac12)$ & & & \\
\hline
\multicolumn{6}{c}{\texttt{vacS12} (control)} \\
\hline
0.26 & $-47.619364$ & $-47.619146$ & $-0.22$ & & \\
0.28 & $-47.552687$ & $-47.554449$ & $+1.76$ & & \\
0.30 & $-47.527245$ & $-47.533793$ & $+6.55$ & & \\
0.32 & $-47.544571$ & $-47.557420$ & $+12.85$ & & \\
\hline\hline
\end{tabular}
\end{center}
\caption{Ground-state energies of the two lowest total-spin sectors of the
doped chain at $L=80$ (bond dimension 2400), and their difference $\Delta$.
The ground state is the lower of the two; the sector is the label the sweep
is carried out in, not a quantity inferred afterwards. The two couplings
closest to the crossing, $0.31$ and $0.315$, were added to bracket it. For
\texttt{vacS1} the last two columns give the singlet--triplet gap of the clean
chain at the same $J_2$ and $L$ and the binding energy
$E_b=\Delta_{\rm ST}-\Delta$, with sign; its size dependence, at $J_2=0.28$
and $0.30$, is given in Sec.~\ref{sec:defect} and vanishes as $1/L$.}
\label{tab:sectors}
\end{table*}

\begin{figure}[tbp]
\centering
\includegraphics[width=\linewidth]{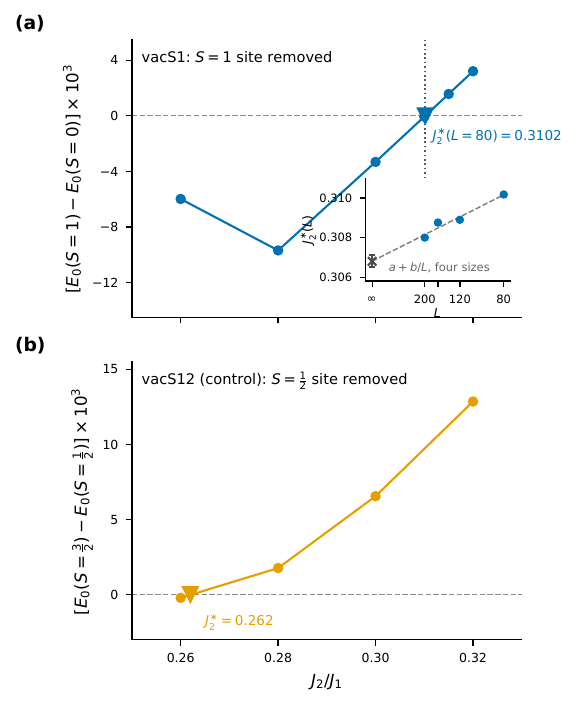}
\caption{Sector splitting of the doped chain at $L=80$. (a) \texttt{vacS1}:
$E_0(S{=}1)-E_0(S{=}0)$ against $J_2$; the triangle marks the sign change,
$J_2^{\ast}=0.3102$, and the dotted line carries it to the top of the panel.
Negative means the doped chain has a triplet ground state. Inset: the crossing
against inverse size, located at each $L$ by the same pair $(0.30,0.32)$ for
$L=80$, $120$, $160$ and $200$; the dashed line is the linear fit in $1/L$
through the four, $J_2^{\ast}(\infty)=0.3068(3)$, Eq.~(\ref{eq:jstarinf});
the point at $L=160$ carries the largest residual. (b) \texttt{vacS12}, the
control: the analogous splitting crosses zero near $J_2=0.262$ instead.}
\label{fig:gap}
\end{figure}

\emph{The ground-state sector of the doped chain switches at $J_2^{\ast}$.}
For \texttt{vacS1} the $S=1$ sector lies below $S=0$ at $J_2=0.26$, $0.28$ and
$0.30$, and above it at $0.32$; the splitting
$\Delta=E_0(S{=}1)-E_0(S{=}0)$ changes sign between $0.31$ and $0.315$
(Fig.~\ref{fig:gap}a). Linear interpolation on that bracketing pair puts the
crossing at
\begin{equation}
  J_2^{\ast}(L{=}80)=0.3102,
  \label{eq:jstar}
\end{equation}
and a quadratic through the three largest couplings returns the same value to
four digits. This is the same transition the $S^z$-resolved U(1) runs
placed at $0.312(2)$, now read off a symmetry label rather than from a
splitting of nearly degenerate $S^z$ levels: below $J_2^{\ast}$ the ground
state of the doped chain is a genuine spin-1 multiplet, above it a singlet.
The splitting is not monotonic --- it deepens from $-5.99\times10^{-3}$ at
$0.26$ to $-9.69\times10^{-3}$ at $0.28$ before rising through zero --- so a
linear form describes only the neighborhood of the crossing. Inside that
neighborhood it describes it well: the four couplings $0.30$, $0.31$, $0.315$
and $0.32$ lie on a straight line to within $3\times10^{-6}$,
$\Delta=0.326\,(J_2-J_2^{\ast})$, and the two couplings that bracket the
crossing most closely, $0.31$ and $0.315$, put it between them. Three
independent interpolations --- the pairs $(0.30,0.32)$, $(0.31,0.315)$ and
$(0.31,0.32)$ --- agree to $2\times10^{-5}$, so at this size
$J_2^{\ast}(L{=}80)=0.31019$ with a fit residual far below the last digit.

\emph{The crossing moves with size as $1/L$.} That number is a finite-size
quantity, and it is worth saying so plainly. Repeating the sector comparison
at $L=120$, bracketing couplings included, gives $J_2^{\ast}(L{=}120)=0.30886$,
lower by $1.3\times10^{-3}$; there too the three interpolations --- the pairs
$(0.30,0.32)$, $(0.31,0.315)$ and $(0.31,0.32)$ --- agree, to $1\times10^{-4}$,
and the four couplings lie on a line to $1\times10^{-5}$ with a slope $0.203$
against $0.326$ at $L=80$, the splitting itself shrinking with size. At
$L=160$ and $200$ the two sectors were converged at $J_2=0.28$, $0.30$ and
$0.32$, and the pair $(0.30,0.32)$, the one protocol available at every size,
gives $J_2^{\ast}(L)=0.31018$, $0.30890$, $0.30876$ and $0.30800$ at $L=80$,
$120$, $160$ and $200$.

The four values lie on a straight line in $1/L$,
$J_2^{\ast}(L)=J_2^{\ast}(\infty)+b/L$ with $b=0.27$, residuals within
$\pm3\times10^{-4}$, and extrapolate to
\begin{equation}
  J_2^{\ast}(\infty)=0.3068(3),
  \label{eq:jstarinf}
\end{equation}
where the uncertainty is the intercept's from the scatter of the four points;
a quadratic term in $1/L$ moves it by $10^{-4}$, and dropping any one size by
at most $3\times10^{-4}$; the first two sizes alone give $0.3063$. The small
step from $L=120$ to $160$ ($1.4\times10^{-4}$) is the largest residual of
the line ($+2.8\times10^{-4}$ at $L=160$), not a saturation of the drift.
The crossing is a property of the scaled spectrum. Near $J_2^{\ast}$ the
slope of the splitting, $0.326$ at $L=80$ and $0.203$ at $L=120$, falls as
$1/L$, so that $L\Delta$ crosses zero with a slope nearly independent of size
($26.1$ and $24.4$). Both levels approach the ground state as $L\to\infty$,
as they must in a critical host; what $J_2^{\ast}$ marks is a change in the
ordering of the lowest levels of the finite-size spectrum in units of $1/L$,
that is, in the boundary condition the vacancy imposes on the host.
What the drift does not touch is the
separation that this section is about: at every size the $S=1$ vacancy
crosses far from where the $S=\tfrac12$ control does, and the host has no
feature at either coupling.

\emph{The control transitions elsewhere.} Removing an $S=\tfrac12$ site
instead leaves a half-integer chain whose two lowest sectors cross at
$J_2\simeq0.262$ at $L=80$ (Fig.~\ref{fig:gap}b), far from $J_2^{\ast}$ and
close to the first-order boundary $J_2^{c1}$. Its size dependence is not a
drift but an alternation: at $L=160$ the two sectors are again degenerate at
$J_2=0.26$ (splitting $-3\times10^{-5}$, against $+4.4\times10^{-4}$ at
$0.28$), while at $L=120$ the splitting is already $+4.3\times10^{-3}$ at
$0.26$. The control crosses near $0.26$ when $L/8$ is even ($L=80$, $160$)
and not when it is odd ($L=120$), the same parity that modulates the
splitting of the $S=1$ vacancy below. The two defects therefore do not share a
transition: what happens at $0.3102$ is specific to removing the $S=1$ site,
not a property of doping the chain in general.

\emph{The triplet is not bound to the vacancy.} A spin-1 ground state does not
by itself mean that the vacancy binds a moment, and two measurements show that
it does not. The first is where the moment sits. The local moment
$\mu_i=\sum_j\langle\mathbf S_i\!\cdot\!\mathbf S_j\rangle/[S(S+1)]$, which
sums to one in the triplet, places $99.6$ to $99.8\%$ of it on one of the two
segments the vacancy leaves, at every size and at both $J_2=0.28$ and $0.30$,
spread along the whole segment with the period-eight modulation of the host.
The eight sites around the vacancy carry $0.30$, $0.22$, $0.17$ and $0.145$ of
it at $L=80$, $120$, $160$ and $200$ ($J_2=0.28$): the fraction falls as $1/L$
($L$ times it is $24$, $26$, $28$ and $29$) instead of saturating, as it would
for a bound state. At $L=80$ the profile is unchanged, to $6\times10^{-9}$,
when the sweep starts from a different random state, so the choice of segment
is not an accident of the algorithm. Which segment hosts the triplet
alternates with size --- the one to the right of the vacancy at $L=80$ and
$160$, the left one at $120$ and $200$ --- following the parity of $L/8$; the
triplet of the clean chain, for comparison, spreads over both halves. The
second measurement is the energy. A bound moment would keep the binding energy
$E_b=\Delta_{\rm ST}-\Delta$, the clean-chain gap of Table~\ref{tab:gapST}
minus the splitting of the doped chain at the same $J_2$ and $L$, finite as
$L\to\infty$. Instead $E_b\times10^{3}=18.8$, $11.8$, $9.2$ and $7.0$ at
$J_2=0.28$ and $18.2$, $11.9$, $9.1$ and $7.2$ at $0.30$, for $L=80$ to $200$:
$LE_b$ stays between $1.41$ and $1.51$, with the same parity alternation, and
its extrapolation in $1/L$ is within $1\times10^{-3}$ of zero for every fit
form we tried. Below $J_2^{\ast}$ the doped chain is a triplet of the critical
host confined to one segment, and $J_2^{\ast}$ is the coupling at which that
triplet crosses the singlet: a level crossing, not the ionization of a bound
moment. We keep calling $J_2<J_2^{\ast}$ the moment side, meaning the side on
which the doped chain carries $S=1$.

\emph{$J_2^{\ast}$ depends on where the vacancy sits.} Every size in this work
removes the site $v=L/2+1$. Moving the vacancy at $L=80$
(Table~\ref{tab:position}) shows that the crossing is tied to that choice.
When the segment to the left of the vacancy holds an even number of unit
cells ($v=21$, $41$, $61$), the doped chain behaves as described above --- a
triplet ground state below a crossing, the moment on one segment --- with
$J_2^{\ast}=0.3117$, $0.3102$ and $0.3088$; across positions the crossing
moves by $2.9\times10^{-3}$, more than its finite-size drift from $L=80$ to
$200$. When the left segment holds an odd number of cells ($v=11$, $31$, $51$,
$71$) there is no crossing in $[0.30,0.32]$: the ground state is a singlet
throughout, the triplet lies $2$--$3\times10^{-3}$ above it at $v=11$ and $51$
and only $4\times10^{-5}$ to $3\times10^{-4}$ above it at $v=31$ and $71$, and
its moment is shared equally by the two segments. The rule holds at all seven
positions, and we have no explanation for it; a test of the natural
candidate, a dimerization in pairs of cells, on clean chains with an odd
number of cells ($L=78$, $82$) found gaps of the same order as at $L=80$ and
rules it out. The rule survives at $L=160$: with the vacancy at $v=41$
($20$ cells to its left) the crossing is at $J_2^{\ast}=0.30809$ and the
moment sits on one segment, while at $v=51$ ($25$ cells) the singlet is the
ground state at both $0.30$ and $0.32$ and the triplet's moment is shared
equally. At $L=160$ the crossing for $v=41$ lies $6.7\times10^{-4}$ below the
one for the central vacancy, against a spread of $1.5$--$2.9\times10^{-3}$
among positions at $L=80$, so the dependence on position shrinks roughly as
$1/L$ (from two sizes, $L=80$ and $160$). All sizes used for
Eq.~(\ref{eq:jstarinf}) have an even left segment,
so the series $J_2^{\ast}(L)$ is homogeneous; $0.3068(3)$ is strictly the
limit for a vacancy at the center, but the shrinking spread suggests that the
limit does not depend on where an even-cell vacancy sits.

\emph{The bridge is what makes the crossing.} Removing the $J_2$ bond between
$v-1$ and $v+1$ as well as the site cuts the chain into two independent
segments, since the other bonds through $v$ go with it. Without the bridge
the splitting no longer crosses once: at $L=80$ it is $-18.2$, $+24.6$ and
$+3.2\times10^{-3}$ at $J_2=0.28$, $0.30$ and $0.32$, and at $L=120$ it
changes sign twice in the same window ($-5.9$, $+1.8$, $-2.2\times10^{-3}$),
following the separate ground-state spins of the two segments. The single,
linear crossing at $J_2^{\ast}$ is a property of the segments coupled through
the bridge, not of either segment alone.

\begin{table}[tbp]
\begin{center}
\begin{tabular}{ccccccc}
\hline\hline
$v$ & left cells & \multicolumn{3}{c}{$\Delta\times10^{3}$ at $J_2=$} & moment left/right & $J_2^{\ast}$ \\
 & & $0.30$ & $0.31$ & $0.32$ & & \\
\hline
11 & 5  & $+2.87$ & $+2.96$ & $+3.06$ & $0.50/0.50$ & --- \\
21 & 10 & $-6.18$ & $-0.89$ & $+4.44$ & $0.99/0.01$ & $0.3117$ \\
31 & 15 & $+0.04$ & $+0.09$ & $+0.16$ & $0.50/0.50$ & --- \\
41 & 20 & $-3.32$ & $-0.07$ & $+3.20$ & $0.00/1.00$ & $0.3102$ \\
51 & 25 & $+1.92$ & $+1.98$ & $+2.04$ & $0.50/0.50$ & --- \\
61 & 30 & $-1.81$ & $+0.24$ & $+2.26$ & $1.00/0.00$ & $0.3088$ \\
71 & 35 & $+0.05$ & $+0.16$ & $+0.31$ & $0.50/0.50$ & --- \\
\hline\hline
\end{tabular}
\end{center}
\caption{Position of the vacancy at $L=80$: removed site $v$, number of unit
cells in the segment to its left, splitting $\Delta=E_0(S{=}1)-E_0(S{=}0)$ of
\texttt{vacS1}, fraction of the triplet's moment on each segment, and the
crossing where there is one. The configuration used everywhere else in this
work is $v=41$.}
\label{tab:position}
\end{table}

\subsection{Transmission through the vacancy}

At $L=80$ the transmission of Eq.~(\ref{eq:T}), built from the invariant and
measured in the ground multiplet, falls steeply as $J_2$ approaches
$J_2^{\ast}$ from below: $T_{\rm IC}=0.340(20)$ at $J_2=0.26$, $0.082(8)$ at
$0.28$ and $0.018(5)$ at $0.30$; just above the transition, in the singlet
sector at $J_2=0.32$, it is back to $0.086(2)$. The vacancy is thus most
opaque at this size at the coupling just below the crossing.

\begin{figure*}[t]
\centering
\includegraphics[width=0.88\textwidth]{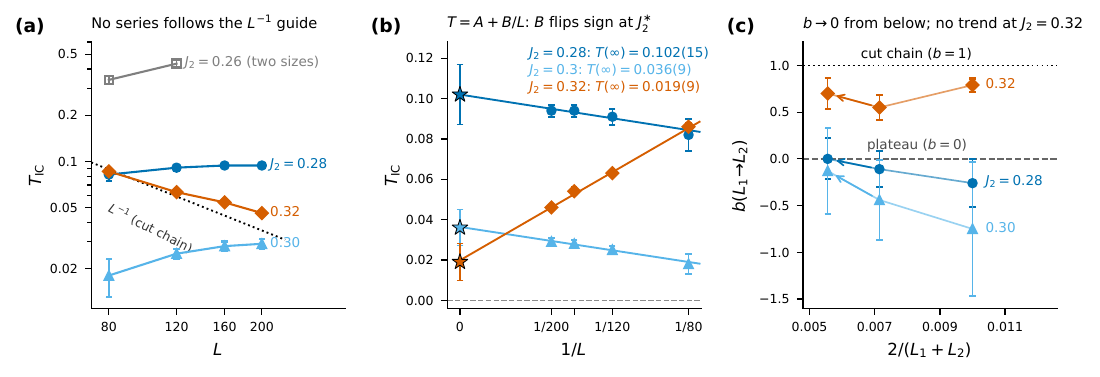}
\caption{Transmission through the $S=1$ vacancy against system size. (a)
Log-log, with the $L^{-1}$ guide of the cut-chain fixed point; open symbols
mark the coupling measured at two sizes only; couplings on the moment side
($J_2=0.28$, $0.30$) are drawn in blue, the singlet side ($0.32$) in orange. No
filled curve tracks the
guide. (b) The same data against $1/L$, with the weighted fit
$T_{\rm IC}=A+B/L$; the stars are the intercepts, $0.102(15)$, $0.036(9)$
and $0.019(9)$ at $J_2=0.28$, $0.30$ and $0.32$ (window systematic included);
the first two are away from the origin that a cut chain requires, the third
is not resolved from a power-law decay. The correction $B$ changes sign with
the sector. (c) The pairwise slope against the inverse mean size; the arrows
run toward larger $L$. On the moment side, at both couplings, the slope
stays negative and reaches $b=0$ from below; on the singlet side it scatters
about $0.6$--$0.7$ with no trend, approaching neither $b=1$ nor $b=0$ within
these sizes.}
\label{fig:TvsL}
\end{figure*}

\emph{Size dependence.} The criterion is the slope
$b(L_1\!\to\!L_2)$, defined as $-\ln(T_2/T_1)/\ln(L_2/L_1)$: $b\to1$ is the
cut-chain fixed point, $b\to0$ a plateau. The slope is negative or zero below
the defect transition and positive above it at every pair of sizes
(Table~\ref{tab:bslope}, Fig.~\ref{fig:TvsL}). Below $J_2^{\ast}$, on
the moment side, the transmission \emph{grows} with $L$ and saturates:
$0.082(8)\to0.091(4)\to0.094(3)\to0.094(3)$ at $J_2=0.28$, i.e.\
$b=-0.26(26)$, $-0.11(19)$ and $0.00(22)$, and
$0.018(5)\to0.025(2)\to0.028(2)\to0.029(2)$ at $0.30$, $b=-0.75(72)$,
$-0.44(43)$ and $-0.13(46)$. At both couplings the slope reaches zero from
below, and at $0.28$ the two largest sizes coincide to the third digit; at
$0.30$ the second slope is negative in every one of the $r$ windows described
below. Above $J_2^{\ast}$, where the vacancy is a singlet, the transmission
decays, $0.086(2)\to0.063(1)\to0.054(2)\to0.046(1)$ at $J_2=0.32$, and its
slope does not settle: $b=+0.79(7)$, $+0.55(13)$ and $+0.70(17)$, scattered
about $0.6$--$0.7$ with no trend, approaching neither $b=1$ nor $b=0$.

Both sides admit a plateau with a $1/L$ correction, $T_{\rm IC}(L)=A+B/L$.
On the moment side it is the physical result: $A=0.102(15)$ and $0.036(9)$
at $J_2=0.28$ and $0.30$, fits with $\chi^2=0.3$ and $0.1$ for two degrees
of freedom, and a pure power law $T\propto L^{-b}$ fitted to the same four
points returns an exponent compatible with zero, $b=-0.04(17)$ and
$-0.28(38)$. The quoted uncertainties are the spread of $A$ over the
estimator's own choices --- thirty-five windows of $r$ with lower edge
$0.08$--$0.12\,L$ and upper edge $0.19$--$0.25\,L$, the same fraction of $L$
at every size --- which exceeds the statistical error of the fit; over that
set the intercept from the two largest
sizes alone stays within $[0.089,0.104]$ at $0.28$ and $[0.029,0.038]$ at
$0.30$. Fixing $q_{\rm IC}$ at $3\pi/4$ instead of the clean peak changes
nothing at the fourth digit. On the singlet side the same fit gives
$A=0.019(9)$ with $\chi^2=1.1$, but a pure power law with $b=0.62(18)$ fits
as well ($\chi^2=1.3$), and the intercept from the two largest sizes alone
ranges over $[0.008,0.023]$ across the windows. Sizes up to $L=200$
therefore do not decide whether the singlet-side transmission goes to a
small plateau or to zero as a power law; we report both and claim neither.
What the data do exclude, on both sides, is the cut chain with the measured
exponent $\eta=1$ (Appendix~\ref{app:scaling}): $T\propto L^{-1}$ gives
$\chi^2=215$, $69$ and $92$ for three degrees of
freedom at $0.28$, $0.30$ and $0.32$, and its logarithmic variant
$T=A\,L^{-1}(\ln L)^{p}$ needs $p=1.5(2)$ at $0.32$ ($\chi^2=3.4$ against
$1.1$ for the plateau) and $p>5$ on the moment side, an exponent with no
predicted value. The correction $B$ changes sign with the sector. Over a
factor of $2.5$ in $L$, then, the transmission does not follow the cut-chain
law on either side of $J_2^{\ast}$; on the moment side it saturates; and
what $J_2^{\ast}$ marks sharply is the sign of the leading finite-size
correction. The surviving transmission falls across the
moment side, $0.10\to0.036$ between $0.28$ and $0.30$, before $J_2^{\ast}$;
whether any survives on the singlet side is the question these sizes leave
open.

The entanglement entropy across the vacancy, $S_{\rm cut}$ (the cut between
$v-1$ and $v+1$), tells the same story from the side of the wave function and
adds one check. In the $S=0$ sector, measured at four sizes at $J_2=0.28$ and
$0.30$, it does not grow with the system: $S_{\rm cut}=0.076$, $0.070$,
$0.069$ and $0.067$ at $L=80$, $120$, $160$ and $200$ for $J_2=0.30$, a fit
$a+\kappa\ln L$ giving $\kappa=-0.010$ (and $\kappa=-0.06$, noisier, at
$0.28$), whereas the clean chain cut at the same bond grows, from $1.66$ to
$2.06$ between $L=80$ and $200$ at $J_2=0.30$. The vacancy therefore carries
almost no entanglement between the two segments, whatever the transmission
of the correlator. In the $S=1$ sector $S_{\rm cut}$ is not a smooth function
of $L$ at all: it is $0.07$--$0.08$ at $L=80$ and $160$ and $1.155$ at $L=120$
and $200$, at both couplings, and the difference, $1.07$--$1.09$, is $\ln 3$ to within
a few percent. That is the entropy $\ln(2s+1)$ of the spin-1 multiplet that the
SU(2) bond carries when the triplet sits on the left segment,
present or absent according to the parity of $L/8$ --- the same alternation
that decides on which segment the triplet sits (Sec.~\ref{sec:defect}), seen
here in an independent observable. A coefficient of $\ln L$ is therefore
defined only in the singlet sector, where it vanishes.

\begin{table*}[t]
\begin{center}
\footnotesize\setlength{\tabcolsep}{5pt}
\begin{tabular}{lcccccccc}
\hline\hline
$J_2$ & $T_{\rm IC}(80)$ & $T_{\rm IC}(120)$ & $T_{\rm IC}(160)$ & $T_{\rm IC}(200)$ &
$b(80\!\to\!120)$ & $b(120\!\to\!160)$ & $b(160\!\to\!200)$ & $A=T_{\rm IC}(\infty)$ \\
\hline
0.26 & 0.340(20) & 0.434(23) & ---       & ---       & $-0.60(19)$ & ---         & ---         & ---       \\
0.28 & 0.082(8)  & 0.091(4)  & 0.094(3)  & 0.094(3)  & $-0.26(26)$ & $-0.11(19)$ & $0.00(22)$  & 0.102(15) \\
0.30 & 0.018(5)  & 0.025(2)  & 0.028(2)  & 0.029(2)  & $-0.75(72)$ & $-0.44(43)$ & $-0.13(46)$ & 0.036(9)  \\
0.32 & 0.086(2)  & 0.063(1)  & 0.054(2)  & 0.046(1)  & $+0.79(7)$  & $+0.55(13)$ & $+0.70(17)$ & 0.019(9)  \\
\hline\hline
\end{tabular}
\end{center}
\caption{Transmission through the $S=1$ vacancy, its pairwise slope
$b(L_1\!\to\!L_2)=-\ln(T_2/T_1)/\ln(L_2/L_1)$, and the intercept $A$ of the
weighted fit $T_{\rm IC}=A+B/L$ to the four sizes, its uncertainty the
spread over the estimator's $r$ windows (Sec.~\ref{sec:defect}). The ground
sector is $S=1$ for $J_2\le0.30$ and $S=0$ at $0.32$, with the crossing at
$J_2^{\ast}(\infty)=0.3068(3)$ (Sec.~\ref{sec:defect}). Cut-chain flow
predicts $b=1$ and $A=0$ (Appendix~\ref{app:scaling}); a plateau, $b=0$ and
$A>0$. The row at $J_2=0.26$, next to the first-order boundary, has two sizes
and no intercept. At $J_2=0.32$ a pure power law $T\propto L^{-0.62(18)}$
fits the four sizes as well as the plateau does.}
\label{tab:bslope}
\end{table*}

The control defect behaves differently again: with a half-integer chain the
$\pi$ component dominates the correlator across the vacancy, the
incommensurate amplitude is a small residue on top of it, and the resulting
$T_{\rm IC}$ carries errors of the order of the value itself. We therefore do
not report slopes for \texttt{vacS12}; it enters this work only through its
sector map (Sec.~\ref{sec:defect}).

\section{Discussion}
\label{sec:discussion}

The singlet phase of the frustrated $(1,\tfrac12)$ chain is entered by a
direct collapse of the total spin, $S=L/4\to0$, between $J_2=0.22$ and $0.24$
at $L=80$, with no partially polarized window: every sector was computed at
$J_2=0.22$ and $0.24$, on the two sides of the transition, and the energy is
monotonic in $S$ at each (inside the narrow coexistence window the low-spin
sectors do not converge, App.~\ref{app:maxwell}); the
sector label of the
SU(2) code leaves no room for an intermediate $S$ to hide in an inferred
$\langle\mathbf S^2\rangle$. Inside it the modulation is commensurate: the
thermodynamic pitch, from four system sizes at $J_2=0.28$, $0.30$ and $0.32$, is
$3\pi/4$ to within $1.1\times10^{-4}$ in the linear extrapolation (a few
$10^{-4}$ allowing for curvature, Sec.~\ref{sec:pitch}) and equal at the three
couplings to $5\times10^{-5}$ (three sizes extend the same picture to
$[0.27,0.32]$), and it neither drifts with $J_2$ nor responds to the defect
transition at $J_2^{\ast}$. The
finite-size pitch does drift, monotonically, and on the two smallest sizes its
$1/L$ amplitude passes through zero near $J_2^{\ast}$; a study at one or two
sizes can read that drift as a moving thermodynamic pitch, but with three and
four sizes the crossing is a property of the correction, not of the limit. This
answers the question left open by
Ref.~\cite{IvanovRichterSchollwock1998}: the phase above the first-order
boundary is a period-eight commensurate singlet, rigid against $J_2$, and
gapless: its singlet--triplet gap closes as $1/L$ over the four sizes, and its
entanglement grows with a central charge $c\simeq1.8$, above the $c=1$ of a
single SU(2)$_1$ chain, while its spin correlations keep the SU(2)$_1$
exponent $\eta=1$.

The vacancy, by contrast, does change its nature inside this rigid host. Read
off the sector label, the doped chain is a spin-1 multiplet for
$J_2\le0.30$ and a singlet at $0.32$, with the level crossing at
$J_2^{\ast}=0.3102$ at $L=80$ and $0.3068(3)$ in the limit; the control
defect, an $S=\tfrac12$ vacancy, crosses elsewhere, near $0.262$, so the
transition at $J_2^{\ast}$ belongs to the removed $S=1$ site and not to doping
in general. What switches is the ordering of the lowest scaled levels, not a
bound moment: the triplet below
$J_2^{\ast}$ is spread over one segment and its binding energy vanishes as
$1/L$, and the crossing exists only for the positions that leave an even
number of cells to the left of the vacancy, moves with the position by an amount that
shrinks with size, and disappears when the bridge across the vacancy is
removed. The transmitted incommensurate
correlations follow the sector, and this is the sharpest statement the defect
data support. With four sizes
the transmission is finite in the limit on the moment side --- it grows with
$L$ and saturates, $T(\infty)=0.102(15)$ and $0.036(9)$ at $J_2=0.28$ and
$0.30$ --- so below $J_2^{\ast}$ the vacancy is not the weak link that the
chain flows away from. On the singlet side it keeps falling, with a slope
near $0.6$--$0.7$ that four sizes drive neither to zero nor to unity: a small
plateau, $0.019(9)$, and a power-law decay to zero are indistinguishable at
$L\le200$, and we do not choose between them. What the transition does
sharply is to reverse the sign of the leading finite-size correction, which
is what made two sizes look like two different fixed points. The reading the
two-size data invited, a cut chain above $J_2^{\ast}$ and a plateau below,
survives in half: the plateau below is established; the cut chain above is
excluded in its pure form, $T\propto L^{-1}$, but not the possibility of a
slower decay to zero. The triplet sector keeps the channel open, then;
whether the singlet closes it or only narrows it is what a fifth size would
have to say. That the defect can switch while the host shows no feature in
its thermodynamic pitch is the central point: the two transitions are decided
by different degrees of freedom. The host does not register $J_2^{\ast}$
anywhere that survives to the limit. It appears to, in the sign of the
finite-size correction to the pitch read on $L=80$ and $120$; the larger
sizes show that this sign change moves with the pair of sizes used, from
$0.3076(43)$ on $(80,120)$ to $0.3263(18)$ on $(120,160)$ and back below
$0.30$ on $(160,200)$, where the correction is no longer resolved from zero
(Sec.~\ref{sec:pitch}). The apparent
signature and its withdrawal are the same lesson the transmission teaches: at
these sizes the smaller pairs are not yet asymptotic, and a quantity
built from two sizes can carry a feature that a third erases.

On method, the result of Sec.~\ref{sec:invariant} has a consequence beyond
this model. In any sector with $S\neq0$ the longitudinal correlator
$\langle S^z_iS^z_j\rangle$ depends on which member of the multiplet the
sweep happens to converge to, and the dependence is not small: it is the
difference between a transmission of $0.055$ and one of $0.005$ at the same
coupling. Transmissions measured with $C^{zz}$ in a U(1) code, in a sector
that turns out to be a triplet, are therefore measurements of a
U(1)-breaking choice made by the algorithm, not of the defect. The
invariant $\langle\mathbf S_i\!\cdot\!\mathbf S_j\rangle$, or an SU(2) code
in which the sector is a label, is the only safe route whenever the defect
carries a moment.

\emph{Experimental signatures.} The separation of the two transitions is
what a measurement could test. Any probe coupling to the local spectral
weight across a vacancy --- NMR relaxation at the sites adjacent to a defect,
or end-state spectroscopy in spin chains assembled atom by
atom~\cite{Huang25,Mishra21} --- should see the low-lying triplet of the
doped chain appear and disappear as the frustration is tuned through the
moment region, with its weight on one segment only (which one is set by the
parity of the segment lengths), while the
bulk modulation, accessible to diffraction or to bulk NMR, evolves smoothly
and stays commensurate. The bimetallic realizations of the
$(1,\tfrac12)$ chain~\cite{Kahn1987,Hagiwara1998} have fixed $J_2$; the
engineered chains do not, and it is there that the switch could be observed.

\section{Conclusion}
\label{sec:conclusion}

Using an SU(2)-symmetric DMRG in which the total spin is a sector label, we
have shown that the frustrated alternating spin-$(1,\tfrac12)$ chain leaves
its ferrimagnetic phase by a direct collapse from $S=L/4$ to $S=0$, and that
the singlet phase it enters carries a modulation locked at the commensurate
wavevector $3\pi/4$, within $1.1\times10^{-4}$ in the linear extrapolation
at the three couplings measured at four sizes, $J_2=0.28$, $0.30$ and $0.32$,
and is gapless, its singlet--triplet gap closing as $1/L$ and its entanglement
entropy growing with a central charge $c\simeq1.8$, while the spin
correlations decay with the SU(2)$_1$ exponent $\eta=1$.
Inside this rigid host a single $S=1$ vacancy switches the ground state of
the doped chain from a triplet to a singlet at $J_2^{\ast}=0.3102$ at $L=80$,
$0.3068(3)$ in the limit for a vacancy at the center, while the host shows no
feature at that coupling. The triplet is not bound to the vacancy --- it
spreads over one segment and its binding energy vanishes as $1/L$ --- and the
crossing depends on the position of the vacancy, existing only when the
segment to its left holds an even number of unit cells. The transmission
through it saturates on the moment side, $T(\infty)=0.102(15)$ and
$0.036(9)$ at $J_2=0.28$ and $0.30$, and on the singlet side keeps falling
with a slope near $0.6$--$0.7$ that sizes up to $L=200$ cannot resolve into a
plateau or a decay to zero; on both sides the cut-chain law $T\propto L^{-1}$
is excluded, and the sign of the finite-size correction is the transition's
sharp signature. The host registers this coupling in no quantity that
survives to the thermodynamic limit: a sign change in the finite-size
correction to its pitch, close to $J_2^{\ast}$ on the two smaller sizes,
moves with the pair of sizes and vanishes from resolution on the largest.
Finally, the longitudinal correlator is
not an observable in a triplet sector, so transmissions through a defect that
carries a moment must be built from the SU(2) invariant.

\section*{Acknowledgements}
A.~R.~P. thanks INCT/CNPq -- Spintr\^onica e Nanoestruturas Magn\'eticas
Avan\c{c}adas (INCT-SpinNanoMag) and Rede Mineira de Nanomagnetismo/FAPEMIG.

\paragraph{Author contributions}
R.~L.~S. designed the study, wrote the SU(2) DMRG code and carried out the
calculations; R.~L.~S., R.~C.~S. and A.~R.~P. analyzed the results and wrote the
manuscript.

\paragraph{Funding information}
This work was supported by the Brazilian agencies CNPq, FAPEMIG, CAPES and
FAPES.

\paragraph{Data and code availability}
The converged energies, sector maps, correlators and entanglement profiles
behind Tables~1--7, the
scripts that produce Figs.~1--4 from them, and the SU(2) DMRG driver
(\textsc{TensorKit} and \textsc{MPSKit}, Julia) are deposited at 10.5281/zenodo.23248228.

\appendix
\numberwithin{equation}{section}

\section{First-order boundary: Maxwell construction and lever rule}
\label{app:maxwell}

\begin{figure*}[t]
\centering
\includegraphics[width=0.88\textwidth]{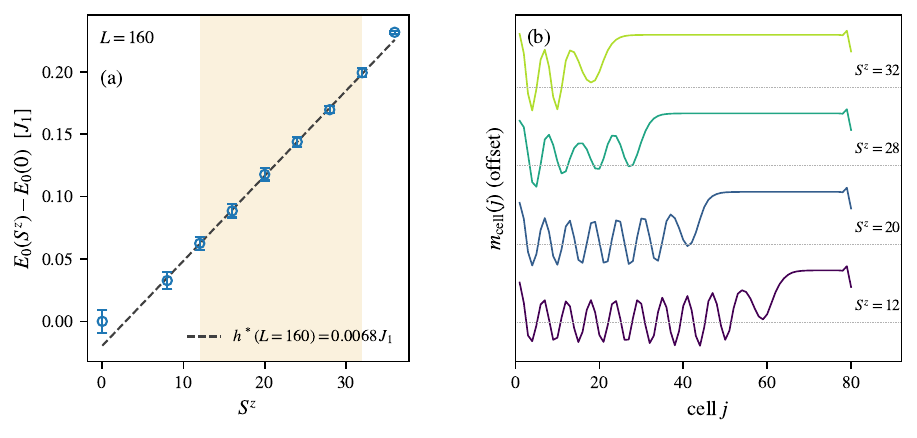}
\caption{Lever-rule test at $J_2/J_1=0.24$, $L=160$ (U(1) code). (a) Sector
energy $E_0(S^z)-E_0(0)$ against $S^z$; the linear regime (shaded) defines the
coexistence field $h^{\ast}$. (b) Cell magnetization
$m_{\rm cell}(j)=\langle S^z_{A,j}\rangle+\langle S^z_{B,j}\rangle$ for
selected $S^z$ (curves offset vertically). The ferrimagnetic domain grows at
the expense of the incommensurate one as $S^z$ increases; the single domain
wall moves linearly with $S^z$.}
\label{fig:maxwell}
\end{figure*}

The location $J_2^{c1}=0.233(3)$ quoted in Sec.~\ref{sec:bulk} comes from an
$S^z$-resolved study of the same chain with a U(1) code, of which we keep
here only the part that certifies the order of the transition. If the
transition is first order, constraining the total magnetization to an
intermediate value must force the chain to phase-separate. At $J_2=0.24$ and
$L=160$ the sector energy $E_0(S^z)$ is linear over a wide range of $S^z$
[Fig.~\ref{fig:maxwell}(a)], as two-phase coexistence requires: each added
unit of magnetization displaces an interface at a fixed energy cost, the
coexistence field $h^{\ast}$. The real-space profiles
[Fig.~\ref{fig:maxwell}(b)] show the mechanism: a single domain wall
separates a ferrimagnetic domain (cell magnetization $\simeq\tfrac12$) from an
incommensurate one ($\simeq0$), and its position advances linearly with
$S^z$. Fitting the number of ferrimagnetic cells gives
$N_{\rm ferri}\simeq1.96\,S^z$ (correlation coefficient $0.994$), against the
lever-rule prediction $N_{\rm ferri}=2S^z$ for phases of cell magnetization
$0$ and $\tfrac12$. The Maxwell slope decreases monotonically with size,
$h^{\ast}=0.00751(34)$, $0.00709(14)$, $0.00707(11)$ and $0.00681(7)\,J_1$
at $L=80$, $100$, $120$ and $160$, and extrapolates linearly in $1/L$ to
$h^{\ast}(\infty)=0.0062(2)\,J_1$. Just above $J_2^{c1}$ a DMRG run
initialized in the ferrimagnetic state converges to that branch even though
it is no longer the ground state (at $J_2=0.235$, $L=60$, it sits
$\simeq0.03\,J_1$ above it), the expected hysteresis of a first-order
transition; such runs are detected by a variational guard and recomputed.

The SU(2) code brackets the transition at $L=80$ without the U(1) data. At
$J_2=0.225$ the ferrimagnetic sector is the ground state, $E_0(S\!=\!20)$ lying
$1.4\times10^{-3}J_1$ below $E_0(S\!=\!0)$; at $J_2=0.23$ it is no longer, the
sectors $S=8$, $12$ and $14$ lying $4.0\times10^{-4}$, $2.7\times10^{-5}$ and
$1.7\times10^{-5}J_1$ below $S=20$, so that $J_2^{c1}(L\!=\!80)$ falls between
$0.225$ and $0.23$, just below the U(1) estimate $0.233(3)$; the boundary drifts
with $L$ at the level of these differences.
That intermediate sectors lie below $S=20$ at $0.23$ is what coexistence
implies at fixed $S$ and does not signal a partially polarized phase: inside
the window $E(S)$ is nearly flat, as the U(1) Maxwell construction above
shows, and the margins between these sectors are too small to resolve an
interior minimum from a coexistence plateau at the accuracy of these states.
Inside the coexistence window the SU(2) runs meet the
difficulty that the U(1) hysteresis already signals: the intermediate
sectors converge only to an energy variance of order $10^{-4}$, and the
low-spin sectors at $J_2=0.23$ and $0.235$ (including $S=0$) do not converge
within three hours at bond dimension $800$, nor does $J_2=0.23$--$0.24$ at
$L=120$. We therefore do not use SU(2) energies inside the window, and the
order of the transition rests on the lever-rule test above.

\section{Scaling of the transmission observable at the cut-chain fixed point}
\label{app:scaling}

$T_{\rm IC}$ of Eq.~(\ref{eq:T}) is a ratio of correlator amplitudes, not a
single-particle transmission probability, and its finite-size exponent must
be derived for the correlator itself. We do so by power counting at the
cut-chain fixed point, to first order in the residual tunneling between the
two half chains.

\emph{Scaling dimensions.} At the cut-chain fixed point the system is two
decoupled semi-infinite Luttinger liquids, each with an open boundary at the
vacancy, coupled by
$H_\lambda=\lambda\,(O^{\dagger}_{B,l}O_{B,r}+{\rm h.c.})$, where $O_{B}$ is
the boundary transverse-spin operator of each half. The bulk transverse
correlator $\langle S^+_0S^-_r\rangle\sim\cos(qr)/r^{\eta}$, $\eta=1/(2K)$,
measured as $\eta=1.0(1)$ in Sec.~\ref{sec:pitch}, fixes the bulk dimension
$x_{\rm bulk}=\eta/2=1/(4K)$; at an open boundary the
density field obeys a Dirichlet condition and, by the method of images, the
boundary counterpart carries twice the bulk dimension,
$x_B=2x_{\rm bulk}=1/(2K)$~\cite{EggertAffleck1992}. At $\eta=1$ ($K=1/2$) this gives
$x_B=1$, the known dimension of the end spin of the open Heisenberg chain, a
consistency check of the convention. The tunneling operator then has boundary
dimension $x_\lambda=2x_B=1/K$, so $d\lambda/d\ell=(1-1/K)\lambda$: the weak
link is irrelevant for all $K<1$, as in the Kane--Fisher
problem~\cite{KaneFisher1992b}, and at $K=1/2$, $\lambda(L)\sim\lambda_0L^{-1}$.

\emph{First-order correlator across the defect.} At $\lambda=0$ the
transmitted correlator vanishes, since the two factors of
$C_{\rm def}(r)=\langle\mathbf S_{v-r}\!\cdot\!\mathbf S_{v+r}\rangle$ live on
decoupled chains. The leading term is first order in $\lambda$,
\begin{align}
C_{\rm def}(r)={}&-\lambda\!\int\! d\tau\,
\langle S^+_{v-r}(0)\,O_{B,l}(\tau)\rangle_{L}\nonumber\\
&\times\langle O^{\dagger}_{B,r}(\tau)\,S^-_{v+r}(0)\rangle_{R}+{\rm c.c.},
\label{eq:Cdef}
\end{align}
each factor a boundary-to-bulk correlator of a semi-infinite critical system,
\begin{equation*}
\langle O_{\rm bulk}(r,0)O_B(0,\tau)\rangle\sim
r^{-(x_{\rm bulk}-x_B)}(r^2+\tau^2)^{-x_B}.
\end{equation*}
The integral over $\tau$ converges for
$4x_B>1$ and is dominated by $|\tau|\lesssim r$, which gives
$C_{\rm def}(r)\sim\lambda\,r^{\,1-2x_{\rm bulk}-2x_B}$. Normalizing by the
clean chain, $C_{\rm clean}(r)\sim r^{-2x_{\rm bulk}}$, and evaluating at fixed
$r/L$ (the windows of Sec.~\ref{sec:estimators}),
\begin{equation}
T_{\rm IC}(L)\sim\lambda_0\,L^{\,1-2x_B}=\lambda_0\,L^{-(1/K-1)},
\label{eq:Tscaling}
\end{equation}
i.e.\ the amplitude ratio tracks the renormalized tunneling. In terms of the
measured exponent, $b=2x_B-1=2\eta-1$, and $\eta=1$ gives $b=1$. Identifying $T_{\rm IC}$ with an
$S$-matrix probability would give $2(1/K-1)=2$, twice the correct value: the
observable is linear, not quadratic, in the tunneling.

\emph{Scope.} The derivation treats the spin channel as a single Luttinger
liquid. The central charge $c\simeq1.8$ of Sec.~\ref{sec:pitch} shows that
the host has a second gapless mode; it does not enter the leading spin
correlator, whose exponent $\eta=1$ is what fixes $b$, but a boundary
operator of that mode could add a channel across the vacancy with a
different exponent. The prediction $b=1$ is therefore the cut-chain behavior
of the spin channel, which is the channel $T_{\rm IC}$ measures.

\emph{Logarithmic corrections.} If the spin channel is an SU(2)$_1$ theory,
as $\eta=1$ suggests, it hosts the marginally
irrelevant operator $\mathbf J_L\!\cdot\!\mathbf J_R$, whose running coupling
$g(\ell)\sim1/\ell$ feeds into the flow of $\lambda$ through operator mixing.
The generic outcome is a multiplicative logarithm,
$T_{\rm IC}(L)=A\,L^{-1}[\ln(L/L_0)]^{p}$, so that the pairwise slopes behave
as $b_{\rm eff}(L)\simeq1-p/\ln(L/L_0)$: whatever the sign of $p$, the
magnitude of the correction decreases with $L$ and $b_{\rm eff}$ approaches
$b=1$ monotonically. This makes the class falsifiable by a few sizes. On
the moment side it is falsified outright: a negative slope is outside it
altogether, whatever $p$. On the singlet side the measured slope,
$+0.79(7)$, $+0.55(13)$ and $+0.70(17)$ (Sec.~\ref{sec:defect}), neither
approaches unity monotonically nor moves away from it --- it scatters about
$0.6$--$0.7$ --- so the form is not excluded there by the direction of $b$,
only disfavored by its fit ($p=1.5(2)$ as a free parameter,
$\chi^2=3.4$ against $1.1$ for a plateau). The derivation above therefore
stands as the prediction for a defect that does flow to the cut chain, and
as the reference this one is measured against --- established as not
describing the moment side, and not decided on the singlet side.

\bibliographystyle{apsrev4-2}
\bibliography{ferri_vacancy_SciPost}

\end{document}